**RESEARCH**

# Pop Music Highlighter: Marking the Emotion Keypoints

Yu-Siang Huang*,†, Szu-Yu Chou*,† and Yi-Hsuan Yang*

The goal of music highlight extraction, or thumbnailing, is to extract a short consecutive segment of a piece of music that is somehow representative of the whole piece. In a previous work, we introduced an attention-based convolutional recurrent neural network that uses music emotion classification as a surrogate task for music highlight extraction, assuming that the most emotional part of a song usually corresponds to the highlight. This paper extends our previous work in the following two aspects. First, methodology-wise we experiment with a new architecture that does not need any recurrent layers, making the training process faster. Moreover, we compare a late-fusion variant and an early-fusion variant to study which one better exploits the attention mechanism. Second, we conduct and report an extensive set of experiments comparing the proposed attention-based methods to a heuristic energy-based method, a structural repetition-based method, and three other simple feature-based methods, respectively. Due to the lack of public-domain labeled data for highlight extraction, following our previous work we use the RWC-Pop 100-song data set to evaluate how the detected highlights overlap with any chorus sections of the songs. The experiments demonstrate superior effectiveness of our methods over the competing methods. For reproducibility, we share the code and the pre-trained model at https://github.com/remyhuang/pop-music-highlighter/.

**Keywords:** Music thumbnailing; highlight extraction; chorus detection; structure analysis; convolutional neural network; attention mechanism

## Introduction

With the growing amount of multimedia data available on the Internet, the ability to efficiently browse this data is important. Music highlight extraction, or music thumbnailing, is such a task that aims to find a short, continuous segment of a piece of music that can nicely represent the whole piece. It can be understood as an audio preview chosen by machines. Successful algorithms for music highlight extraction are useful for many music information retrieval (MIR) tasks, such as indexing, retrieval (Lee et al., 2014), trial listening (Goto, 2003), radio podcasts (Mehrabi et al., 2017) and DJing (Bittner et al., 2017). For example, we can use highlight extraction as a pre-processing step to pick a representative segment from each song to facilitate the subsequent labeling, processing or analysis, instead of dealing with the whole song or taking a random segment per song (e.g., the middle 30 seconds).

The selection of the highlighted music piece is a subjective matter and different people may have different opinions. However, for Pop/Rock music, there is a stronger agreement that the chorus (refrain) section of a song is a representative sample of the whole song.[1] It is a common practice to play the chorus when introducing a Pop music hit chart or when using music in commercials (Goto, 2003). A user study conducted by Meintanis and Shipman III (2008) showed that, for getting familiar with a song, the chorus section was regarded as the most important part of a song, compared with other parts such as the intro or the verse. Some MIR researchers view chorus sections as the most memorable and emotional part of a song (Goto, 2006; Eronen, 2007). By extracting features only from the chorus (instead of the verse or the whole song), Wang et al. (2013) demonstrated improved accuracy in recognizing song level arousal- and dominance-related emotions.

### Related Work

Due to the absence of well-defined and public domain labeled data for music highlights, an important stream of related work focuses on chorus detection instead. Many previous researchers assumed that the most-repeated patterns or melody motifs correspond to the chorus sections. They firstly used methods such as the self-similarity matrix (SSM) (Cooper & Foote, 2003; Bartsch and Wakefield, 2005) or hidden Markov model (HMM) (Logan & Chu, 2000; Peeters et al., 2002) for segmenting a music piece into several parts, and then analyzed (e.g., by clustering) the resulting segments to identify the most frequent ones. Commonly used feature representations include time-frequency representations such as the spectrogram and the chromagram (Müller and Ewert, 2011). Because the chorus repeats several times in a song, it is still necessary to employ some methods to pick one of them as the highlight. A simple solution is to randomly pick one, if no other heuristics are used. We see two limitations in such an approach. First, the most-repeated part of a song may not always correspond to the chorus; it can also be the verse or even other short melodies. Second, there are still some variations between different chorus sections of a song (such as key change (Goto, 2006)) and it is better to give these sections different "highlight scores" to distinguish them. In other words, a classifier or a regression model is needed. However, little work has been performed regarding this, to the best of our knowledge, again due to the lack of labeled data for training a supervised model that estimates such highlight scores.

* Research Center of IT Innovation, Academia Sinica, TW
† Graduate Institute of Networking and Multimedia, National Taiwan University, TW
Corresponding author: Yu-Siang Huang (yshuang@citi.sinica.edu.tw)

Another important stream of related work attempts to detect highlights by leveraging traces of user behavior logged by online music services. For example, working on electronic dance music (EDM), Yadati et al. (2014) aimed to detect *drops*, the specific moments where there is drastic change in a song. Such drops are usually associated with strong emotional release. Because a drop point often implies the most interesting part of an EDM song, we can take the segment starting from a drop point as the highlight. Yadati et al. achieved drop detection by feeding the spectrogram, Mel-Frequency Cepstral Coefficients (MFCC) and features related to rhythm to a support vector machine (SVM). The labeled data needed to train the SVM were mined from SoundCloud,[2] which allows users to leave "timed comments" annotating specific moments (e.g., the drop point) of a song. This idea seems to work for EDM, but for other music genres it is unclear whether such timed comments are available.

Another approach utilizing user logs was presented by Bittner et al. (2017). It has been pointed out[3] that people may "scrub" (i.e., move) the playhead of a music player to start a song at a specific moment, and such moments usually occur just before the best parts of the song.

Therefore, if one has access to scrubbing behavior data as Bittner et al. did, highlight extraction can be performed by simply running a peak picking algorithm over the scrubbing data to find the key moments. In this *crowd-sourced* method, no classifiers are needed. Streaming companies such as Spotify[4] can anonymously log such scrubbing behavior data. However, to our knowledge such data are not yet open to the research community. Moreover, a classifier may still be needed if we want to have the highlight of unpopular songs (i.e., songs in the "long tail").

We took a different approach in a previous work (Huang et al., 2017b) by utilizing the possible connections between emotions and song highlights. The idea is to firstly use a data set with emotion labels to train a neural network model for music emotion classification (Yang and Liu, 2013). Then, it adds on top of the network a so-called "attention mechanism" to weigh the contribution of different short time chunks of a song (e.g., each 3 seconds in length) in predicting the song-level emotion labels of that song. The attention mechanism has been widely used in many natural language processing (NLP) problems (Bahdanau et al., 2014; Shen et al., 2017; Gehring et al., 2017; Vaswani et al., 2017) to learn to "pay attention" to specific parts of an input. We used an attention mechanism to assign "attention scores" (i.e., weights) to audio chunks to facilitate a weighted aggregation of the emotion prediction from different chunks in making the final song-level prediction, as illustrated in **Figure 1(a)**. While the model was trained using emotion labels only (i.e., not using any annotations of the chorus sections or the highlights), we investigated whether peaks of the attention scores help detect the chorus sections of songs. A preliminary experiment showed that this idea is promising (Huang et al., 2017b).

In parallel to our work, Ha et al. (2017) independently proposed a similar attention-based neural network model for music highlight extraction. There are three notable differences between their model and ours. First, they used music genre classification as the surrogate task, while we used emotion classification. Their attention mechanism assessed how each audio chunk contributes to predicting the genre of a song. Second, while we made an emotion prediction per audio chunk first and then used attention to arrive at the song-level estimate (i.e., making decision-level, or *late fusion* (Atrey et al., 2010)), they used attention to linearly combine the features extracted from each chunk and then performed song-level prediction based on the aggregated feature vector (i.e., feature-level, or *early fusion* (Atrey et al., 2010)), as depicted in **Figure 1(b)**. Lastly, while we used attention scores alone for highlight extraction, they additionally considered the energy curve of the input audio. This makes sense because it has been found that chorus-like sections and dramatic moments of a song tend to be louder than other parts of the same song (van Balen et al., 2013).[5] However, Ha et al. (2017) evaluated the result of their model for highlight extraction using a private data set (i.e., highlight sections labeled by human annotators hired by NAVER Music[6]) and they did not report the result without using the energy. Moreover, while combining the energy curve and the attention scores, they actually weighed the energy curve more (0.9 versus 0.1). Therefore, it can be argued that their highlights were determined mainly by energy.

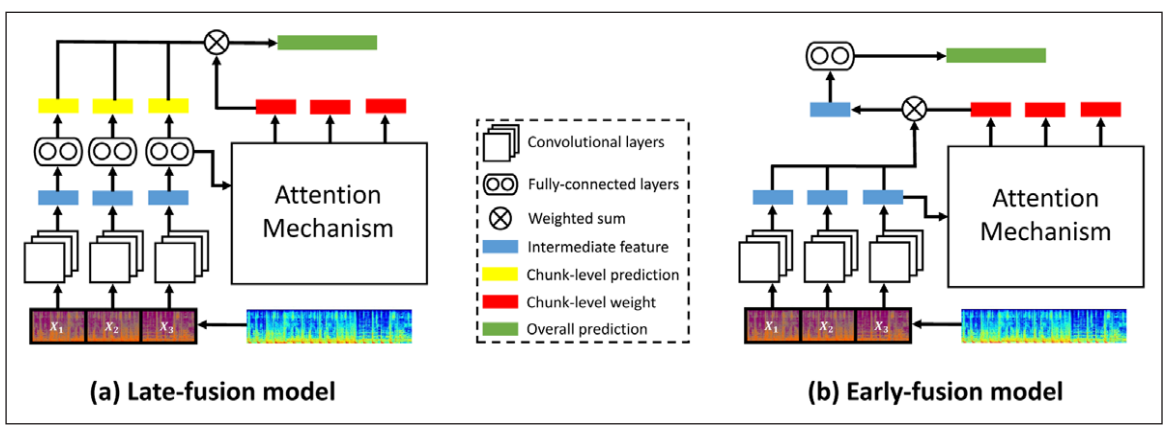

**Figure 1:** Architecture of two attention-based models using different fusion methods for highlight extraction. We note that model **(a)** was used by Huang et al. (2017b) and model **(b)** by Ha et al. (2017).

*Our Contribution*

The goal of this paper is to extend and complement our previous work (Huang et al., 2017b) and the work presented by Ha et al. (2017) in the following aspects:

- We use a public-domain chorus detection data set, the RWC-Pop database (Goto et al., 2002), to evaluate the effect of the following two main design choices of the attention-based approach for highlight extraction: 1) using emotion classification or genre classification as the surrogate task; 2) using late fusion or early fusion to incorporate attention scores. These two facets represent the main difference between our model and the model proposed by Ha et al. (2017).
- It has been shown by Ha et al. (2017) that it is possible to speed up the model training process by using convolutional layers rather than recurrent layers for the attention mechanism, at moderate performance drop in highlight extraction. Motivated by the work of Vaswani et al. (2017), we study in this paper a more advanced design that uses the so-called positional encodings (Gehring et al., 2017) to model the temporal ordering of audio chunks and uses only fully-connected layers for the attention mechanism. We observe a slight boost in the accuracy of chorus detection with the new design.
- While we only compared with a repetition-based method (Nieto and Bello, 2016) in our previous work, we consider here three simple unsupervised feature-based methods for performance comparison, covering the use of spectral energy, spectral centroid and spectral roll-off. Moreover, motivated by Ha et al. (2017), we investigate the performance of fusing the energy-based method and attention-based methods.[7]

Music highlight extraction has not received much academic attention in the literature. To promote future research and to facilitate application of our model to other MIR problems, we share with the research community the code and the emotion-based highlight extraction model we built at https://github.com/remyhuang/pop-music-highlighter/.[8]

In what follows, we give details of the proposed model and its variants in Section 2, and then report the experiments in Section 3. Before concluding the paper, we discuss limitations of the present work in Section 4.

## Methods

*Recurrent Neural Attention Modeling by Late Fusion (RNAM-LF)*

We firstly introduce the model proposed in our previous work (Huang et al., 2017b), of which the model architecture is illustrated in **Figure 1(a)**. The model uses multi-class music emotion classification as the surrogate task (i.e., the goal of the "overall prediction" in **Figure 1(a)**). In addition, it incorporates a recurrent attention mechanism in a late-fusion fashion, by weighing the chunk-level predictions. Therefore, we call it emotion-based recurrent neural attention modeling by late fusion, or emotion-based RNAM-LF for short.

Specifically, given a song of arbitrary length, we transform the raw audio into its mel-spectrogram and then split the whole input into several chunks with fixed short length. This leads to $T$ chunks of mel-spectrograms, $\mathbf{X}_t$, $t \in \{1, \ldots, T\}$, each of which can be considered a 2-D image. For each chunk, we use a stack of convolutional layers (*Conv*) for high-level feature extraction, followed by a global max pooling layer over the temporal dimension for temporal aggregation (Liu and Yang, 2016; Choi et al., 2017). This way, we convert each 2-D chunk input into a 1-D feature vector $\mathbf{h}_t \in \mathbb{R}^M$:

$$\mathbf{h}_t = \text{TimeMaxPool}(\text{Conv}(\mathbf{X}_t)). \quad (1)$$

After that, we pass the intermediate features $\mathbf{h}_t$ to two different branches. One is for the prediction outputs, which uses fully-connected layers (*FC*) to learn the mapping between the intermediate features and target emotion labels. As a result, each chunk has its own chunk-level label prediction $\hat{\mathbf{y}}_t \in \mathbb{R}^C$:

$$\hat{\mathbf{y}}_t = \text{softmax}(\text{FC}(\mathbf{h}_t)), \quad (2)$$

where $C$ denotes the number of classes. The softmax function is used to squash the output of the final *FC* layer in the range [0, 1] that adds up to 1. The other branch is for an attention mechanism that calculates the importance (i.e., attention score) of each chunk.

One significant characteristic of music is that it is sequential, meaning that a musical event is likely to have certain dependency upon its preceding events, and such dependency tends to be stronger if two musical events are nearby. Hence, it is sensible to exploit the temporal information while estimating the attention scores. In RNAM-LF, the attention mechanism is realized by a recurrent layer with the bi-directional[9] long-short term memory (LSTM) cells (Hochreiter and Schmidhuber, 1997). We extract the forward output $\mathbf{f}_t$ and backward output $\mathbf{b}_t$ of the LSTM for every chunk. Next, the attention score $\alpha_t$ for the chunk is computed by:

$$\mathbf{f}_t, \mathbf{b}_t = \text{Bi-LSTM}(\mathbf{h}_t), \tag{3}$$

$$\alpha_t = \text{softmax}(\mathbf{W}_1 \tanh(\mathbf{W}_2 \mathbf{f}_t + \mathbf{W}_3 \mathbf{b}_t + \sigma)), \tag{4}$$

where $\mathbf{W}_1$, $\mathbf{W}_2$, $\mathbf{W}_3$ and $\sigma$ are learnable parameters. We then merge the two branches (see **Figure 1(a)**) by using $\alpha_t$ to linearly combine the chunk-level predictions to arrive at the song-level prediction $\hat{\mathbf{y}} \in \mathbb{R}^C$:

$$\hat{\mathbf{y}} = \sum_{t=1}^{T} \alpha_t \hat{\mathbf{y}}_t. \tag{5}$$

Because the attention scores are used after information from each chunk has been used to make a label prediction, we call this a late-fusion model. In the training stage, the estimate $\hat{\mathbf{y}}$ is compared with the ground truth label $\mathbf{y} \in \{0, 1\}^C$ to compute the binary cross-entropy loss, which is then backpropagated to update the network parameters.

We expect that the attention mechanism can allow the model to weigh different chunks differently, based on the output of the memory cells in the LSTM. Therefore, the final estimate of the model is likely to boost the contribution of the relevant chunks and suppress that of the unimportant ones.

### Non-recurrent Neural Attention Modeling by Late Fusion (NAM-LF)

In the field of NLP, Vaswani et al. (2017) recently proposed a network architecture known as "the Transformer," which contains only attention mechanisms (they called these *self-attention* layers), and uses no convolutional or recurrent layers at all. Besides, to better exploit the temporal information, they added the positional encoding (Gehring et al., 2017) to the input features to represent its relative position in the entire sentence. This new design makes both model training and inference faster (since there is no recurrent layer) and it exhibits strong performance for machine translation tasks.

Motivated by this idea, we experiment with a new design that does not use recurrent layers for the attention mechanism. Specifically, we add the (chunk-level) positional embeddings $\mathbf{p}_t$ to the intermediate feature embeddings $\mathbf{h}_t$, before feeding them to the attention mechanism. The positional embeddings have the same dimension $M$ as $\mathbf{h}_t$. Following (Vaswani et al., 2017), we use sine and cosine functions of different frequencies for the positional encoding:

$$p_{t,2z-1} = \sin(t/10000^{2(z-1)/M}), \tag{6}$$

$$p_{t,2z} = \cos(t/10000^{2(z-1)/M}), \tag{7}$$

where $p_{t,j}$ denotes the $j$-th element of $\mathbf{p}_t$, and $z \in \{1, \ldots, M/2\}$. We can see that $\mathbf{p}_t$ only encodes the positional information of each chunk and it has nothing to do with the actual audio content of the chunk.

Then, we use two fully-connected (*FC*) layers to realize the attention mechanism:

$$\mathbf{h}_t^{(pos)} = \mathbf{h}_t + \mathbf{p}_t, \tag{8}$$

$$\mathbf{f}_t = \text{FC}(\mathbf{h}_t^{(pos)}), \tag{9}$$

$$\alpha_t = \mathbf{W}_2 \tanh(\mathbf{W}_1 \mathbf{f}_t + \sigma_1) + \sigma_2. \tag{10}$$

We then similarly use the attention scores for late fusion. This new model is different from RNAM-LF only in details regarding the attention mechanism block in **Figure 1(a)**. We use "NAM-LF (pos)" to refer to this new model. Moreover, we use "NAM-LF" to refer to the variant that uses $\mathbf{h}_t$ instead of $\mathbf{h}_t^{(pos)}$ as input to the fully-connected layers in Eq. (9), to study the effect of the positional encoding.

NAM-LF and NAM-LF (pos) are computationally lighter than RNAM-LF. Moreover, NAM-LF (pos) explicitly takes into account the position of each chunk when calculating the attention scores. On the one hand, this helps the model extract sequential information from a song. On the other hand, information regarding the position of chunks may be useful on its own right. For example, intuitively, the highlight of a song tends not to appear in the beginning of the song.

### Early-fusion Variant (NAM-EF)

As discussed earlier and also illustrated in **Figure 1(b)**, we consider an early-fusion variant that uses the attention scores to combine the feature embeddings $\mathbf{h}_t$ and then uses the aggregated feature vector $\sum_{t=1}^{T} \alpha_t \mathbf{h}_t$ to make the song-level estimate through a stack of fully-connected layers. We use 'EF' to denote such a model that adopts this early-fusion method.

As shown in **Figure 1**, for either late fusion (LF) or early fusion (EF) the attention mechanism can be viewed as a supplementary network in addition to the main network (i.e., a convolutional neural network here) for making (emotion) predictions. However, in LF the attention score is used at the very end of the whole network, whereas in EF there are more fully-connected layers after the fusion is done. Therefore, the attention mechanism in LF has more "stress" on finding out the importance of chunks, for there are no other learnable parameters after it. Accordingly, the LF design may perform better than EF, when we want to use the estimated attention scores for highlight extraction. We intend to verify this in our experiment.

### Genre-based Variant

We can also use multi-label or multi-class music genre classification as the surrogate task (Ha et al., 2017) to train the models. This amounts to only changing the labeled data; we can use exactly the same network architectures introduced before.

### Music Highlight Extraction

We can now generate a music highlight of pre-defined length based on the attention scores. For the listening experience, the extracted highlight has to be a consecutive segment. Hence, we simply use a running window over the attention curve $\{\alpha_t\}_{t=1}^{T}$ to pick a consecutive collection of chunks with the highest aggregated sum. Although this method is simple, it provides a direct way to assess what the attention mechanism captures.

### Implementation Details

We show the details of the network architecture of NAM-LF (pos) as an example in **Table 1**. All the other attention-based models we implemented and evaluated in our experiments employ almost the same architecture (e.g., the same number of layers, filter size, activation functions (Choi et al., 2017)).

For the input data, we converted the raw audio of each song into mel-scale spectrograms with 128 components using LibROSA (McFee et al., 2015), sampling the song at 22,050 Hz and using a Hamming window of 2,048 samples and a 512-sample hop size. Moreover, following common practice (van den Oord et al., 2013; Dieleman and Schrauwen, 2014) we transformed the mel-scale spectrograms to the log scale by the function $g(x) = \log(1 + 10000x)$.

For emotion labels, we used an in-house data collection compiled by Yang and Liu (2013), consisting of 31,377 clips of Western Pop music with song-level emotion labels crawled from the music service website AllMusic.[10] The songs are from the 1970s to 2000s. Among the 190 possible emotion classes (e.g., 'Happy', 'Sad', 'Aggressive', 'Peaceful'), the most popular emotion is associated with 248 songs, the least popular emotion has 28 songs, and the average number of songs per emotion class is about 165. Instead of having access to the full songs, we only have a 24-second audio preview crawled from 7digital[11], without knowing the starting point where the audio preview was extracted from in each song. We randomly sample 19,377 songs for training the network and 6,000 songs as the validation set.[12] Each clip is uniformly segmented into $T = 8$ chunks with no overlaps, each 3 seconds. Accordingly, the size of each chunk input $\mathbf{X}_t$ is $129 \times 128$.[13]

As shown in **Table 1**, we use three convolution layers for feature extraction, two fully-connected layers for chunk-level prediction, and four fully-connected layers for the attention mechanism. For RNAM-LF, we use one layer of the bi-directional LSTM for the attention mechanism, with 512 hidden units. No matter which model is used, the output for each song $\hat{\mathbf{y}}$ is a vector of length $C = 190$.

For the genre-based variant, we collect a new data set from Beatport[14], a worldwide principal source of music for DJs, for EDM sub-genre classification. It contains in total 30 possible sub-genres of EDM, such as 'House' and 'Dubstep'.[15] Most songs were released between 2014 and 2017. For each class, we collect 700 songs for training and 100 songs for validation, leading to a training data set with 21,000 songs and a validation set with 3,000 songs. Because a song is associated with only one sub-genre, we treat it as a multi-class classification problem. Each audio file we acquire from Beatport is a 2-minute sample version of the song (again starting from an unknown position of the song). We take only the first 24 seconds so that we can use the same network architecture as the emotion-based counterpart.[16]

We use a mini-batch of 16 songs for model training, which amounts to $16 \times 8 = 128$ chunks per mini-batch. We find it useful to apply batch normalization (Ioffe and Szegedy, 2015) at every layer of the network to stabilize the training process. At the inference (testing) stage, we process only one song at a time. However, as we divide a song into 3-second chunks, we can also view these chunks as a mini-batch and apply batch normalization to normalize the chunks based on their shared statistics. We find doing so empirically improves the results of highlight extraction. We note that each test song can be of arbitrary length, so the number of chunks for each test song varies.

**Table 1:** Network architecture of the proposed NAMLF (pos) model. For convolutional layers (conv), the values represent (from left to right in the same row): number of filters, kernel size, strides and activation functions. For fully-connected layers, the values represent: number of hidden units, dropout rate and activation functions. All layers use batch normalization. We also show the size of the input, output and all the intermediate output of the training stage.

| (Assume mini-batch size is 16; each clip has 8 chunks) Input $\in \mathbb{R}^{16 \times 8 \times 129 \times 128}$ | | | | |
|---|---|---|---|---|
| reshape, $\{\mathbf{X}_t\} \in \mathbb{R}^{128 \times 129 \times 128}$ | | | | |
| **Feature extraction** | | | | |
| conv | 64 | 3 × 128 | (2, 128) | ReLU |
| conv | 128 | 4 × 1 | (2, 1) | ReLU |
| conv | 256 | 4 × 1 | (2, 1) | ReLU |
| global max-pool to $\mathbb{R}^{128 \times 256}$ reshape, $\{\mathbf{h}_t\} \in \mathbb{R}^{16 \times 8 \times 256}$ | | | | |
| **Attention mechanism** | | | | |
| add positional encodings $\in \mathbb{R}^{16 \times 8 \times 256}$ | | | | |
| fully-connected | 256 | 0.5 | | ReLU |
| fully-connected | 256 | 0.5 | | ReLU |
| fully-connected | 256 | 0.5 | | tanh |
| fully-connected | 1 | 0.5 | | linear |
| softmax along the second axis $\mathbb{R}^{16 \times 8 \times 1}$ reshape, $\{a_t\} \in \mathbb{R}^{16 \times 8}$ | | | | |
| **Chunk-level prediction** | | | | |
| fully-connected | 1024 | 0.5 | | ReLU |
| fully-connected | 190 | 0.5 | | softmax |
| $\{\hat{\mathbf{y}}_t\} \in \mathbb{R}^{16 \times 8 \times 190}$ | | | | |
| **Song-level prediction** | | | | |
| $\hat{\mathbf{y}} = \hat{\mathbf{y}}_t \odot a_t$ | | | | |
| Output, $\{\hat{\mathbf{y}}\} \in \mathbb{R}^{16 \times 190}$ | | | | |

### Experiment

Due to the lack of public-domain annotations of music highlights, we use chorus detection as a proxy to evaluate the result of highlight extraction for the performance study reported here.

*Evaluation Protocol: Chorus Detection*

We use the 100-song popular music subset of the RWC database (Goto et al., 2002) to evaluate the correspondence between music highlights and chorus sections. In RWC-Pop, 20 songs are in English and have the style of Western Pop music typical of songs on the American hit charts in the 1980s. The rest 80 songs are in Japanese and in the style of Japanese Pop music typical of songs on the Japanese hit charts in the 1990s. RWC is publicly available and it has been widely used in MIR. Accordingly, researchers can easily obtain a copy of the data set to evaluate their highlight extraction model and to fairly compare their results with ours.

The RWC-Pop data set comes with manual annotation of the start and end times of each chorus section in each song. In our experiment, we fix the length of the music highlights to 30 seconds and evaluate the degree of overlap between this 30-second highlight, and the (variable-length) chorus section of the song with the largest overlap with the highlight. This can be quantified by the recall rate ($R$) and the precision rate ($P$) defined as follows (Goto, 2006):

$$R = \frac{\text{the length of the overlap}}{\text{the length of the nearest chorus section}}, \quad (11)$$

$$P = \frac{\text{the length of the overlap}}{\text{the length of the extracted highlight}}, \quad (12)$$

where "the length of the overlap" represents the temporal overlap between the extracted highlight and the nearest chorus section. We also compute the F-measure, $F = \frac{2RP}{R+P}$, the harmonic mean of $R$ and $P$. If the highlight has no overlap with any chorus sections of a song, R, P and F are all zero. If the highlight is a subset of the nearest chorus section, or if the nearest chorus section is a subset of the highlight, R, P, and F will be close to one. We calculate these values for each song in RWC-Pop and then report the average results.

*Methods*

We consider the following attention-based methods:

- **RNAM-LF:** RNAM with late fusion.
- **NAM-LF (pos):** NAM (i.e., no recurrent layer) with late fusion and positional encoding.
- **NAM-EF (pos):** NAM with early fusion and positional encoding.
- **NAM-LF:** NAM with late fusion but no positional encoding.

We consider both the emotion-based and genre-based variants of these models. **Emotion-based RNAM-LF** is in principal the method proposed in our previous work (Huang et al., 2017b), so we simply use our previous implementation. **Genre-based NAM-EF (pos)** is the setting that is closest to the one proposed by Ha et al. (2017), with three differences: 1) they used either recurrent or convolution layers for the attention mechanism, while we use fully-connected layers plus the positional encoding; 2) we use our own genre classification data set; 3) we use our own settings regarding the details of the network (e.g., number of layers) to facilitate fair comparison of our own implementation of these models. Because of these differences, we note that the results of genre-based NAM-EF (pos) reported here cannot reflect the performance of their model (Ha et al., 2017), despite the similarity in the main ideas.

In addition, we consider the following baselines:

- **Upper bound:** As the length of the extracted highlight is fixed to 30 seconds and the length of the chorus sections are variable (mostly shorter than 30 seconds), it is not possible to have 100% R and P for all the songs. The upper bound performance is obtained by an oracle method that selects a 30-second segment per song that leads to the highest R and P.
- **Middle:** Simply takes the middle 30 seconds as the highlight of a song.
- **Spectral energy:** Uses LibROSA to compute the root-mean-square energy for each frame of the spectrogram to create the energy curve, and then chooses the segment with the highest aggregated sum as the highlight.
- **Spectral centroid:** Similar to the last one but uses spectral centroid curve instead, which is related to the perceptual "brightness" of sound (Tzanetakis and Cook, 2002).
- **Spectral roll-off:** The last simple feature-based method. The feature has been used to distinguish unvoiced and voiced frames of audio signals (Tzanetakis and Cook, 2002).
- **Repetition:** An unsupervised method that uses the "structural features" proposed by Serra et al. (2014) for music segmentation and then a 2D-Fourier magnitude coefficient-based clustering method (Nieto and Bello, 2014) for grouping the resulting musical segments. We assume the largest cluster (i.e., the most frequent segment) corresponds to the chorus and choose the segment in the cluster that is closest to the middle of the song as the highlight. For a fair comparison, we trim or extend the segment to make it 30 seconds long. Our implementation is based on the Music Structure Analysis Framework (MSAF) proposed by Nieto and Bello (2016).[17]

*Results*

We tabulate the results in **Table 2**, dividing them into three groups: unsupervised methods, emotion-based attentional modeling methods, and the genre-based counterparts.[18] From the results of the first group, the energy-based method performs the best, reaching 0.7562 F-measure. Moreover, there is a large performance gap between this method and all the other methods. This suggests that the chorus sections tend to be louder than other parts and this can be a useful feature. The results of the first group also show that taking the middle segment or the most repeated one does not work well in finding the chorus sections.

By comparing the results of the first and second groups in **Table 2**, it appears the attention-based method, when using emotion classification as the surrogate task, outperforms the energy-based method. In particular, NAM-LF (pos) leads to the highest F-measure (0.7994), R (0.9017) and P (0.7397).[19] This is encouraging, as it implies that chunks contributing more to the prediction of music emotions may correspond to song highlights, and that such connections can be leveraged by the proposed methods.

The following observations can be made by comparing the results within the second group in **Table 2**: 1) LF works better than EF, 2) the positional encoding is useful for NAM, 3) with the positional encoding, NAM slightly outperforms RNAM. This validates our previous choice of LF over EF (Huang et al., 2017b), and shows that the attention mechanism can be simplified by using fully-connected layers.

In addition to performing better in highlight extraction, discarding the recurrent layers also speeds up the training procedure. In our implementation, the training time of NAM per epoch is about three times shorter than that of RNAM.

**Table 2:** Performance of different music highlight extraction methods for chorus detection.

|  | Method | F-measure | Recall | Precision |
|---|---|---|---|---|
|  | Upper bound | 0.9493 | 0.9997 | 0.9173 |
| Unsupervised | Middle | 0.3558 | 0.4708 | 0.2943 |
|  | Spectral energy | **0.7562** | **0.8608** | **0.6960** |
|  | Spectral centroid | 0.5385 | 0.6285 | 0.4867 |
|  | Spectral roll-off | 0.5080 | 0.6059 | 0.4563 |
|  | Repetition | 0.4795 | 0.5973 | 0.4110 |
| Emotion | RNAM-LF | 0.7803 | 0.9006 | 0.7097 |
|  | NAM-LF (pos) | **0.7994** | **0.9017** | **0.7397** |
|  | NAM-EF (pos) | 0.7686 | 0.8727 | 0.7073 |
|  | NAM-LF | 0.7739 | 0.8760 | 0.7120 |
| Genre | RNAM-LF | **0.6314** | **0.7488** | **0.5663** |
|  | NAM-LF (pos) | 0.5891 | 0.6993 | 0.5273 |
|  | NAM-EF (pos) | 0.4688 | 0.5649 | 0.4167 |
|  | NAM-LF | 0.5685 | 0.6725 | 0.5127 |

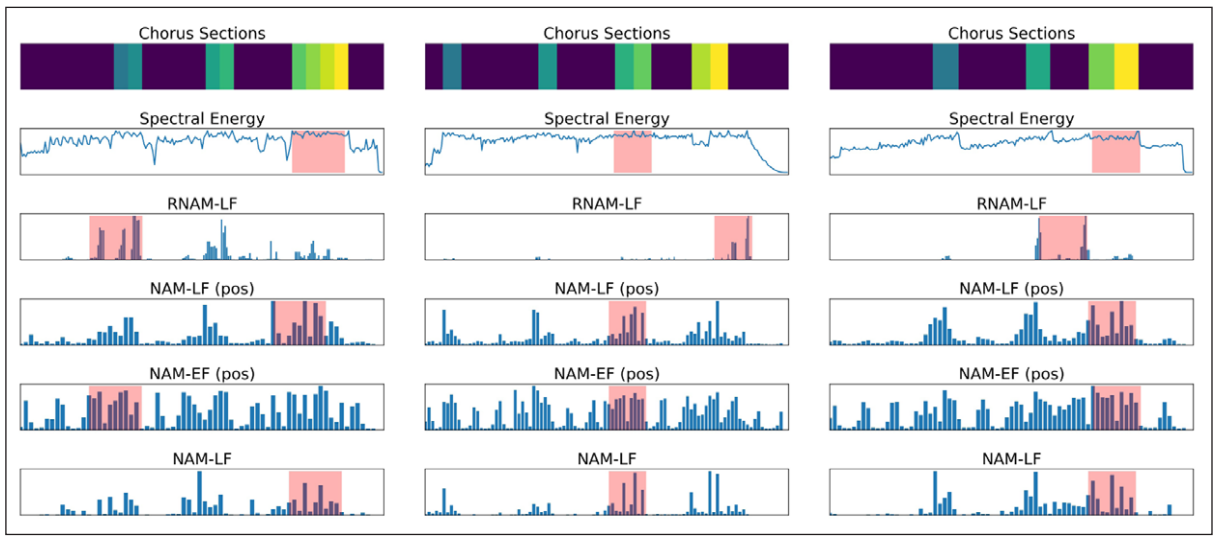

**Figure 2:** Top row: the ground truth chorus sections, where different colors indicate different chorus sections (e.g., chorus A and chorus B) of a song. Second row: the energy curve. Last four rows: the attention curves estimated by four different emotion-based models, for three songs in RWC-Pop. From left to right: 'Disc1/006.mp3', 'Disc2/003.mp3' and 'Disc3/008.mp3'. In RNAM-LF, we have an attention score for each 1-second audio chunk, following our previous work (Huang et al., 2017b); for the other three attention-based methods, we have an attention score for each 3-second audio chunk. The red regions mark the resulting 30-second highlights. More examples can be found on the github page.

**Figure 2** shows the energy curve and attention curves computed by different methods for three randomly selected songs from RWC-Pop, and how the resulting highlights correspond to the chorus sections. We see that different attention-based methods sometimes attend to different choruses of the same song, but the NAM-based method tends to select the same chorus as the energy-based method. Moreover, by comparing the attention curves of NAM-LF (pos) and NAM-EF (pos), the former has fewer peaks, and most of those peaks nicely correspond to different choruses of the same song. This again confirms that LF may be a better choice than EF.

Comparing the results of the second and third groups in **Table 2** shows that the genre-based models are much inferior to the emotion-based models. In the genre-based settings, RNAM-LF outperforms NAM-LF (pos) and many unsupervised methods, but the best results here (RNAM-LF) cannot beat even the energy-based method. While it is possible to use other genre data sets to improve the performance, our current implementation shows that emotions exhibit a stronger connection with the choruses than genres do.

**Figure 3** is the genre-version of **Figure 2** and we see that the attention curves here are not as sparse as the emotion-based counterparts. See for example the results of RNAM-LF.

**Figure 4**, lastly, shows the results when we fuse the energy curves with the attention curves as Ha et al. (2017) did. We try different fusion weights running from 1 (purely energy) to 0 (purely attention). For the emotion-based method NAM-LF (pos), the fusion slightly improves the results, especially for the precision rate. The best result is obtained with a fusion weight of 0.5. A possible reason for the performance gain is that the energy curve is frame-based so it provides a better resolution than the attention curve (chunk-based; every 3 seconds) and this helps fine-tune the result.

In contrast, when fusing energy and genre-based attention, we see from **Figure 4** that the best result is obtained by weighing the energy curve more (at 0.8), which is similar to the weight (i.e., 0.9) used by Ha et al. (2017). This is further evidence that it is better to use emotion rather than genre for the surrogate task of the attention-based models.

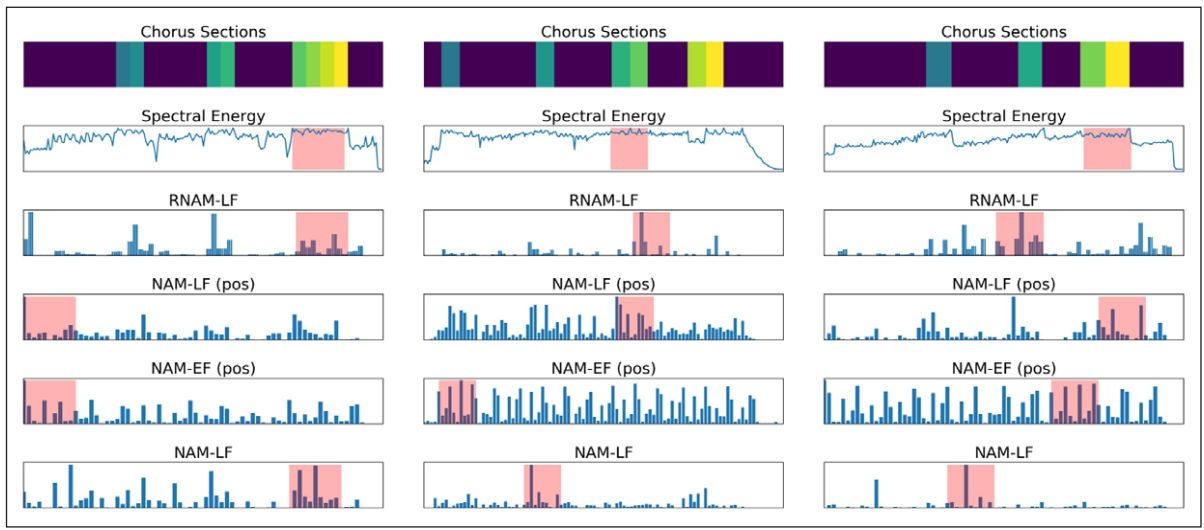

**Figure 3:** Last four rows: Attention curves and the resulting 30-second highlights of different attention-based methods, all genre based, for the same three songs used in Figure 2 (see Figure 2 caption for details).

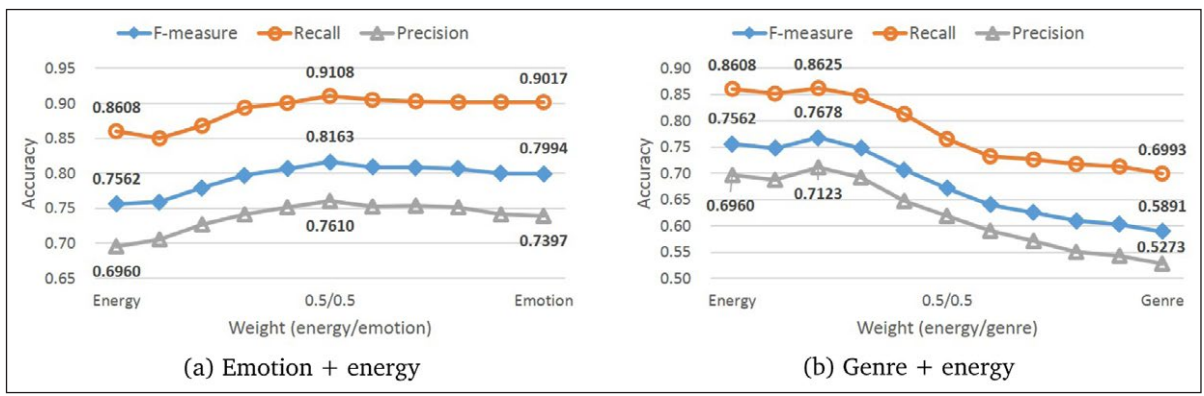

(a) Emotion + energy

(b) Genre + energy

**Figure 4:** Results of chorus detection by fusing the energy curve with the attention curve estimated by either **(a)** emotion-based NAM-LF (pos) or **(b)** genre-based NAM-LF (pos).

## Discussion

In this section, we discuss issues that may have limited the performance of our models. First, we note that the training data we use are 24-second previews of songs. It is possible that such a preview does not contain the highlight or any chorus sections of the song. It would have been better if the training data had consisted of full songs, instead of just previews.

Second, in our current implementation, we use the emotion labels crawled from the All-Music website as the ground truth for training an emotion classifier. It is possible that the labeled data are noisy. In addition, considering up to 190 emotion classes makes it difficult to carefully investigate the dependency between emotions and song highlights. In the future, we hope to validate the emotion labels with human validation (e.g., as Panda et al. (2018) did), and to reduce the number of emotion classes to facilitate in-depth analysis. For example, depending on the emotion a song tries to express, the highlight of the song may not be the chorus but somewhere else.

Moreover, while there are Japanese Pop songs in RWC-Pop, the training data for emotion classification are mainly composed of Western Pop music. On the one hand, we hope to test highlight extraction on other music genres to evaluate generalizability. On the other hand, it may also be important to include songs of other genres in the training set.

As input features to our CNN models, we consider only the mel-spectrogram so far. Other feature representations such as the chromagram (Müller and Ewert, 2011) and the tempogram (Grosche et al., 2010) can be added to exploit information regarding chord progressions or rhythmic patterns.

Finally, we note that the main goal in this work is to extract a continuous short segment from a full-length song as its highlight. Therefore, we do not need to recall all the chorus sections of a song. However, as the chorus sections of a song are similar to one another, it is possible to recall all of them by finding segments that are acoustically similar to the selected highlight. This can be performed for example with an integration of our model and MSAF (Nieto and Bello, 2016).

## Conclusion

In this paper, we have presented a modification of our previous attention-based neural network models that leverage emotions to extract music highlights. This time, we use a non-recurrent attention mechanism inspired by Vaswani et al. (2017). This new design reduces the training time to one third (of the previous approach) and contributes to improved results in highlight extraction. Moreover, we have presented a performance study that provides insights into the design of such an attention-based method, verifying the advantage of using emotion rather than genre for the surrogate task, and of using the late-fusion design rather than the early-fusion design. When evaluating on RWC-Pop for chorus detection, the best method, emotion-based NAM-LF (pos), achieves 0.7994 F-measure, which is better than an energy-based method (0.7562 F-measure) and several other unsupervised baselines. Results shown in **Table 2** indicate that there is still a large performance gap to the oracle method 'Upper Bound' (0.9493 F-measure), but we find the results encouraging as the proposed method does not involve any labels of the chorus sections or music highlights.

For future work, we are interested mainly in the following two directions. First, emotions may not be the only cue for highlights and we want to explore aspects such as aesthetics and novelty for highlight extraction. Second, we hope to build our own or collaborate with companies on a benchmark data set for evaluating music highlight extraction, to avoid the possible discrepancy between choruses and highlights and to help move forward research on this topic.

We hope people will find the "Pop Music Highlighter" model we share on github useful and use it to advance other MIR topics.

## Notes

[1] For sure there are many exceptions. For example, one may consider the legendary guitar solo in the Eagles' *Hotel California* as its highlight.

[2] https://soundcloud.com/.

[3] https://musicmachinery.com/2015/06/16/the-drop-machine/.

[4] https://press.spotify.com/us/about/.

[5] Also see https://musicmachinery.com/2014/09/08/more-on-wheres-the-drama/.

[6] http://music.naver.com/.

[7] It would have been better if we could compare the attention-based method to that presented by Bittner et al. (2017). However, this is not possible without access to the Spotify internal data.

[8] We call our model the "pop music highlighter" because our current implementation considers only pop songs. The extension to other musical genres is left as future work.

[9] The bi-directional design is composed of a forward pass and backward pass that process the data along the temporal dimension.

[10] https://www.allmusic.com/moods.

[11] https://www.7digital.com/.

[12] We do not prepare a test set for this data set and the Beatport data set mentioned later in the paper, as the focus is on evaluating the performance of highlight extraction, not emotion or genre classification.

[13] $\lfloor (3 \times 22050)/512 \rfloor = 129$.

[14] https://www.beatport.com/.

[15] We use the Beatport data set, because EDM sub-genre classification is part of our ongoing DJ-related project (Huang et al., 2017a) and because we have no access to the genre data set used by Ha et al. (2017). The complete list of the 30 subgenres is: 'Breaks', 'Dance', 'Hard Dance', 'House', 'Deep House', 'Electro House', 'Progressive House', 'Dubstep', 'Dub Techno', 'Electro Big Room', 'Electronica/Downtempo', 'Future House', 'Garage/Bassline/Grime', 'Glitch Hop', 'Hardcore/Hard Techno', 'Hip-Hop/R&B', 'Jump Up DnB', 'Jungle DnB', 'Leftfield Bass', 'Leftfield House&Techno', 'Liquid DnB', 'Minimal/Deep Tech', 'Progressive Big Room',

'Progressive Trance', 'Psy Trance', 'Reggae/Dancehall/Dub', 'Tech House', 'Tech Trance', 'Trap/Future Bass', 'Uplifting Trance'.

[16] For the emotion surrogate task, the classification accuracy achieved by NAM-LF (pos) on the validation set is 0.0718, when that of a random guess is 0.0052 (i.e., 1 out of 190). For the genre surrogate task, the classification accuracy achieved by NAM-LF (pos) on the validation set is 0.4763, when that of a random guess is 0.0333 (1/30).

[17] MSAF contains the implementation of various algorithms for music structural analysis. Our selection of the specific algorithms, structural features and 2D-Fourier magnitude coefficient-based clustering, is based on the performance study reported by Nieto and Bello (2016). Via personal communication with Nieto, we learned that unfortunately MSAF does not yet provide a good solution for chorus detection. However, we still include it as a baseline, for it represents a state-of-the-art unsupervised structure analysis method available to the MIR community.

[18] Loosely speaking the attention-based methods may be considered as unsupervised as well for they do not use any labels of chorus sections or highlights.

[19] We note that Precision (P) tends to be lower than Recall (R), as the chorus sections are usually shorter than 30 seconds (see Eqs. (11) and (12)).

## Competing Interests

This work was supported by a grant from the Ministry of Science and Technology of Taiwan under contracts MOST 104-2221-E-001-029-MY3 and 106-3114-E-002-007.